\documentclass[letterpaper,11pt]{article}
\usepackage[margin=1in]{geometry}
\usepackage[utf8]{inputenc}
\usepackage[x11names]{xcolor}
\usepackage{hyperref}
\usepackage{url}
\hypersetup{
    breaklinks,
    colorlinks=true,
    linkcolor=Blue4,
    citecolor=Blue4,
    urlcolor=black,
    pdftitle={Models of High-Level Computation}, pdfauthor={Damian Arellanes} }

\usepackage{amsmath,amsthm,amssymb}
\theoremstyle{definition}

\newtheorem{definition}{Definition}

\usepackage{caption}
\usepackage{subcaption}
\usepackage{wrapfig}
\usepackage{tcolorbox}
\tcbuselibrary{breakable}
\tcbset{
  halign=justify,
  breakable,
  bottomrule=0.5pt,
	toprule=0.5pt,
	leftrule=0.5pt,
	rightrule=0.5pt,
	titlerule=0.5pt,
	opacityfill=0.7,
	standard jigsaw
}
\usepackage{tikz}
\usetikzlibrary{automata,positioning,arrows}
\tikzset{
->, 
>=stealth, 
node distance=1.8cm, 
every state/.style={thick, fill=gray!10}, 
initial text=$ $ 
}

\title{Models of High-Level Computation}
\date{}

\newcommand*\email[1]{\href{mailto:#1}{#1}}
\author{Damian Arellanes \\ Lancaster University \\ \email{damian.arellanes@lancaster.ac.uk}}

\begin{document}

\maketitle

\newenvironment{sideBar}[1][r]
  {\wrapfigure{#1}{0.5\textwidth}\tcolorbox}
  {\endtcolorbox\endwrapfigure}

\begin{abstract}
Classical models of computation have been successful in capturing the very essence of individual computing devices. Although they are useful to understand computability power and limitations in the small, such models are not suitable to study large-scale complex computations. Accordingly, plenty of formalisms have been proposed in the last half century as an attempt to raise the level of abstraction, with the aim of describing not only a single computing device but interactions among a collection of them. In this paper, we encompass such formalisms into a common framework which we refer to as Models of High-Level Computation. We particularly discuss the semantics, some of the key properties, paradigms and future directions of such models.
\end{abstract}

\section{Introduction}
\label{sec:introduction}

Classical models of computation formally emerged during the first half of the century 20th, as an attempt to capture the very essence of computation in the form of information processing. Under the Church-Turing thesis, these models have been successful to describe any possible effective procedure construed as an algorithmic process. Although they can be used to describe complex computations, classical models fail to provide expressive means to do so, apart from not being suitable to accept information streams at computation-time. Describing complex computations has become essential nowadays since the scale and complexity of computing systems is exponentially increasing (especially in the realm of distributed computing). Accordingly, plenty of \emph{Models of High-Level Computation} (MHCs) have been proposed over the last half century so as to capture the essence of complex computations. In this paper, we describe the meaning, key properties, current paradigms and future directions of MHCs. 

The rest of the paper is structured as follows. Section \ref{sec:mhc} describes the general notion of MHCs. Section \ref{sec:example} sketches the design of a simple MHC with the aim of discussing the key elementary properties of such models. Section \ref{sec:classification} presents a preliminary classification of MHCs by considering three major paradigms. Section \ref{sec:conclusions} outlines the conclusions and future directions of the research area this paper introduces.

\section{What is a Model of High-Level Computation?}
\label{sec:mhc}

A MHC raises the level of abstraction of its classical, low-level counterpart (e.g., Turing Machines) by providing a birds-eye-view of multiple computing devices rather than focusing on a single one.\footnote{We should not confuse high-level with higher-order computations \cite{longley_higher-order_2015}. Higher-order computability is a well-established field in theoretical computer science, whose aim is to study computations that receive and produce other computations.} As individual devices are treated as \emph{modular black boxes}, their internal details are irrelevant. What matters is how to glue together multiple (low- or even high-level) computations in order to form more complex ones. Thus, a MHC induces modularity and describes how computing devices interact with the aim of processing information that may be initially encoded, that can continuously be produced by an exogenous entity or any combination thereof.\footnote{By exogenous we mean an entity that is out of the scope of the interacting computing devices, e.g., a human operator.} That is, a MHC can be \emph{open or closed}.\footnote{Classical, low-level models of computation are inherently closed because they have a fixed input before computing (e.g., on a tape) so outputs can only be read after termination. Open models of computation contrast with this behaviour by allowing the processing of data streams from the external world while computing. An example is a process that waits for a human operator's input before proceeding, just as in the original Turing c-machines \cite{turing_computable_1937}.}

By the above, it is evident that a MHC needs to intrisically \emph{separate computation from interaction}. Therefore, a so-called composition mechanism needs to be used. A composition mechanism particularly specifies how computing devices interact from a high-level perspective. An interaction always defines \emph{control flow either implicitly or explicitly}. It is explicit when there is a clear construct specifying the order of invocation of computing devices, and it is implicit otherwise. Apart from control, an interaction can optionally define \emph{implicit or explicit data flow} to establish a data exchange scheme among computing devices.

\begin{tcolorbox}[center,width=\textwidth]
Rather than describing how a single computing device produces an output from a given input, a MHC gives a birds-eye-view on multiple interacting computing devices. For this, a MHC provides formal rules to determine the moment in which a each participant device should compute and how devices must interact. 
\end{tcolorbox}

\begin{figure}[!h]
\centering
\begin{tikzpicture}
\draw[draw=black] (3.8,0) rectangle ++(2.2,2);
\node[align=center] at (4.9,1){Composition\\ Operator};

\node at (0,1.7){(Composite) Computing Device $1$};
\node[label={$\vdots$}] at (0,0.55) {};
\node at (0,0.3){(Composite) Computing Device $M$};
\draw[->] (3,1.7) -- (3.8,1.7);
\draw[->] (3,0.3) -- (3.8,0.3);

\node at (9.4,1){Composite Computing Device};
\draw[->] (6,1) -- (6.8,1);
\end{tikzpicture}         
\caption{Conceptual representation of an algebraic composition operator that takes $M$ computing devices and produces a composite one. Each device operand can be either composite or non-composite.}
\label{fig:composition-operator}
\end{figure}
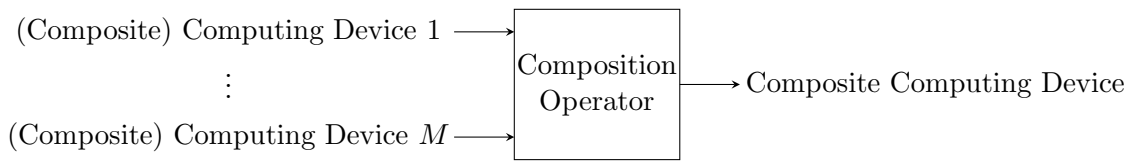

Typically, computing devices facilitate composition by providing \emph{interfaces} that serve as endpoints to entry or exit some internal computational structure. Enabling these endpoints is what allows treating computing devices as high-level, modular black boxes which can be \emph{composed algebraically or non-algebraically}. Only algebraic composition enables \emph{compositionality}, a necessary property to ensure closure by the provision of higher-order operators that receive a number of computing devices and produce a composite one which can, in turn, be used as an operand to construct even more complex computing devices (see Figure \ref{fig:composition-operator}). 

\begin{tcolorbox}[center,width=\textwidth]
A MHC (i) can be open or closed, (ii) induces modularity by black-boxing computing devices (with clear interfaces), (iii) separates computation from interaction, (iv) always defines explicit or implicit control flow, (v) optionally defines implicit or explicit data flow, and (vi) enables interaction among computing devices via a  composition mechanism (which can be algebraic or non-algebraic).
\end{tcolorbox}

\section{Designing a Simple MHC}
\label{sec:example}

For the sake of argument (even if not computationally powerful), in this section we design a simple MHC to demonstrate the key properties discussed in Section \ref{sec:mhc}. For this, we consider computing devices in the form of Nondeterministic Finite Automata (NFAs) as well as composition operators for concatenating ($\oplus$) and parallelising ($\otimes$) --- see Definition \ref{def:nfa}. The behaviour of $\oplus$ and $\otimes$ is directly derivable from the respective proofs of closure under concatenation and under union of regular languages \cite{sipser_introduction_2013}, as shown by Definitions \ref{def:con} and \ref{def:par}. 

\begin{definition}[NFA]\label{def:nfa}
Let $\mathbb{A}$ be the universe of NFAs. A NFA ${N \in \mathbb{A}}$ is a tuple ${(\Sigma,S,s,\delta,F)}$ where $\Sigma$ is a finite set of input symbols which always contains the empty string $\epsilon$, $S$ is a finite set of states, ${s \in S}$ is called an initial state, ${\delta:S \times \Sigma \rightarrow \mathcal{P}(S)}$ is a transition function and ${F \subseteq S}$ is an empty or nonempty set of final states. We say that $N$ accepts a string ${w=x_1 x_2 \ldots x_n}$ over $\Sigma$ if there is a sequence ${(p_0,p_1,\ldots,p_n) \in S^n}$ satisfying the following conditions:
\begin{enumerate}
\item ${p_0=s}$,
\item ${p_{i+1} \in \delta(p_i,x_{i+1})}$ for ${i=0,\ldots,n-1}$ and
\item ${p_n \in F}$.
\end{enumerate}
\end{definition}

\begin{definition}[Concatenative NFA]\label{def:con}
The concatenation operator ${\oplus: \mathbb{A} \times \mathbb{A} \rightarrow \mathbb{A}}$ is a binary function that receives two NFAS, ${N_1=(\Sigma_1,S_1,s_1,\delta_1,F_1)}$ and ${N_2=(\Sigma_2,S_2,s_2,\delta_2,F_2)}$, and produces a concatenative NFA ${N_1 \oplus N_2=(\Sigma_0,S_0,s_0,\delta_0,F_0)}$ where ${\Sigma_0=\Sigma_1\cup\Sigma_2}$, ${S_0=S_1\cup S_2}$, $s_0$ is the initial state $s_1$ of $N_1$, ${F_0=F_2}$ and:
\[
\delta_0(s,x) = 
  \begin{cases}
    \delta_1(s,x) & \text{ if } s \in S_1 \text{ and } s \notin F_1 \\
    \delta_1(s,x) & \text{ if } s \in F_1 \text{ and } x\neq\epsilon \\
    \delta_1(s,x)\cup\{s_2\} & \text{ if } s \in F_1 \text{ and } x=\epsilon \\
    \delta_2(s,x) & \text{ if } s \in S_2 
  \end{cases}
\]
for any state ${s \in S_0}$ and any symbol ${x \in \Sigma_0}$.
\end{definition}

\begin{definition}[Parallel NFA]\label{def:par}
The parallelising operator ${\otimes: \mathbb{A} \times \mathbb{A} \rightarrow \mathbb{A}}$ is a binary function that receives two NFAS, ${N_1=(\Sigma_1,S_1,s_1,\delta_1,F_1)}$ and ${N_2=(\Sigma_2,S_2,s_2,\delta_2,F_2)}$, and produces a parallel NFA ${N_1 \otimes N_2=(\Sigma_0,S_0,s_0,\delta_0,F_0)}$ where ${\Sigma_0=\Sigma_1\cup\Sigma_2}$, ${S_0=S_1\cup S_2 \cup\{s_0\}}$, ${F_0=F_1\cup F_2}$ and:
\[
\delta_0(s,x) = 
  \begin{cases}
    \delta_i(s,x) & \text{ if } s \in S_i \text{ for } i=1,2 \\
    \{s_1,s_2\} & \text{ if } s=s_0 \text{ and } x=\epsilon \\
    \emptyset & \text{ if } s=s_0 \text{ and } x \neq \epsilon
  \end{cases}
\]
for any state ${s \in S_0}$ and any symbol ${x \in \Sigma_0}$.
\end{definition}

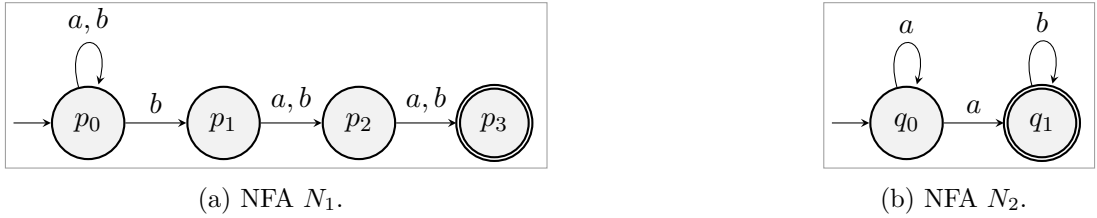
\begin{figure}[!ht]
\centering
\begin{subfigure}[b]{0.6\textwidth}
\centering
\begin{tikzpicture}
\draw[draw=black!38!white] (-1.1,-0.6) rectangle ++(7.2,2.2);
\node[state, initial] (p0) {$p_0$};
\node[state, right of=p0] (p1) {$p_1$};
\node[state, right of=p1] (p2) {$p_2$};
\node[state, accepting, right of=p2] (p3) {$p_3$};
\draw (p0) edge[loop above] node{$a,b$} (p0)
		  (p0) edge[above] node{$b$} (p1)
			(p1) edge[above] node{$a,b$} (p2)
			(p2) edge[above] node{$a,b$} (p3);
\end{tikzpicture}
\caption{NFA $N_1$.}
\label{fig:n1}
\end{subfigure}      
\hfill
\begin{subfigure}[b]{0.3\textwidth}
\centering
\begin{tikzpicture}
\draw[draw=black!38!white] (-1.1,-0.6) rectangle ++(3.6,2.2);

\node[state, initial] (q0) {$q_0$};
\node[state, accepting, right of=q0] (q1) {$q_1$};
\draw (q0) edge[loop above] node{$a$} (q0)
		  (q0) edge[above] node{$a$} (q1)
		  (q1) edge[loop above] node{$b$} (q1);
\end{tikzpicture}         
\caption{NFA $N_2$.}
\label{fig:n2}
\end{subfigure}     
\caption{NFAs serving as building blocks to construct more complex NFAs.}
\label{fig:nas}
\end{figure}

After defining the algebraic composition operators for our MHC, let us now consider the NFA $N_1$ depicted in Figure \ref{fig:n1} which recognises the language ${L_1=\{w\}}$ where $w$ is a string over $\{a,b\}$ containing the symbol ``b" at the third position (from right to left) such as $abaabaa$ \cite{sipser_introduction_2013}. Also, consider the NFA $N_2$ depicted in Figure \ref{fig:n2} which recognises the language ${L_2=\{a^m b^n\}}$ for ${m>0}$ and ${n \geq 0}$.

\begin{figure}[!h]
\centering
\begin{subfigure}[b]{\textwidth}
\centering
\begin{tikzpicture}
\draw[draw=black!38!white] (-1.6,-1.1) rectangle ++(12.2,3.2);

\draw[draw=black!38!white] (-1.1,-0.6) rectangle ++(7.2,2.2);
\node[state, initial] (p0) {$p_0$};
\node[state, right of=p0] (p1) {$p_1$};
\node[state, right of=p1] (p2) {$p_2$};
\node[state, right of=p2] (p3) {$p_3$};
\draw (p0) edge[loop above] node{$a,b$} (p0)
		  (p0) edge[above] node{$b$} (p1)
			(p1) edge[above] node{$a,b$} (p2)
			(p2) edge[above] node{$a,b$} (p3);
						
\draw[draw=black!38!white] (6.8,-0.6) rectangle ++(3.3,2.2);
\node[state, right of=p3, xshift=0.4cm] (q0) {$q_0$};
\node[state, accepting, right of=q0] (q1) {$q_1$};
\draw (q0) edge[loop above] node{$a$} (q0)
		  (q0) edge[above] node{$a$} (q1)
		  (q1) edge[loop above] node{$b$} (q1);
			
\draw (p3) edge[above] node{$\epsilon$} (q0);
\end{tikzpicture}
\caption{Concatenative NFA ${N_1 \oplus N_2}$ (as per Definition \ref{def:con}).}
\label{fig:n1-con-n2}
\end{subfigure}      
\hfill
\begin{subfigure}[b]{\textwidth}
\centering
\begin{tikzpicture}
\node[state, initial] (r0) {$r_0$};
\draw[draw=black!38!white] (-1.1,-2.55) rectangle ++(9.55,5.9);

\draw[draw=black!38!white] (1.2,0.6) rectangle ++(6.7,2.2);
\node[state, right of=r0, yshift=1.2cm] (p0) {$p_0$};
\node[state, right of=p0] (p1) {$p_1$};
\node[state, right of=p1] (p2) {$p_2$};
\node[state, accepting, right of=p2] (p3) {$p_3$};
\draw (p0) edge[loop above] node{$a,b$} (p0)
		  (p0) edge[above] node{$b$} (p1)
			(p1) edge[above] node{$a,b$} (p2)
			(p2) edge[above] node{$a,b$} (p3);

\draw[draw=black!38!white] (1.2,-2) rectangle ++(3.1,2.2);
\node[state, below of=p0, yshift=-0.8cm] (q0) {$q_0$};
\node[state, accepting, right of=q0] (q1) {$q_1$};
\draw (q0) edge[loop above] node{$a$} (q0)
		  (q0) edge[above] node{$a$} (q1)
		  (q1) edge[loop above] node{$b$} (q1);
			
\draw (r0) edge[left] node{$\epsilon$} (p0)
			(r0) edge[left] node{$\epsilon$} (q0);
\end{tikzpicture}         
\caption{Parallel NFA ${N_1 \otimes N_2}$ (as per Definition \ref{def:par}).}
\label{fig:n1-par-n2}
\end{subfigure}     
\caption{Composing the NFAs of Figure \ref{fig:nas} in two different ways. An enclosing box means that the internal details of an automaton are abstracted away. Only initial and final states are visible from a high-level perspective. That is, both $N_1$ and $N_2$ are treated as black boxes.}
\label{fig:composites}
\end{figure}
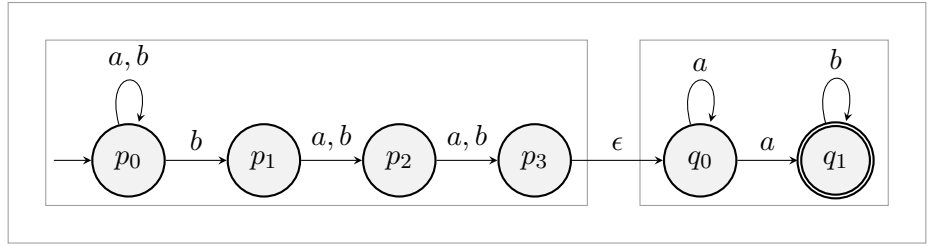
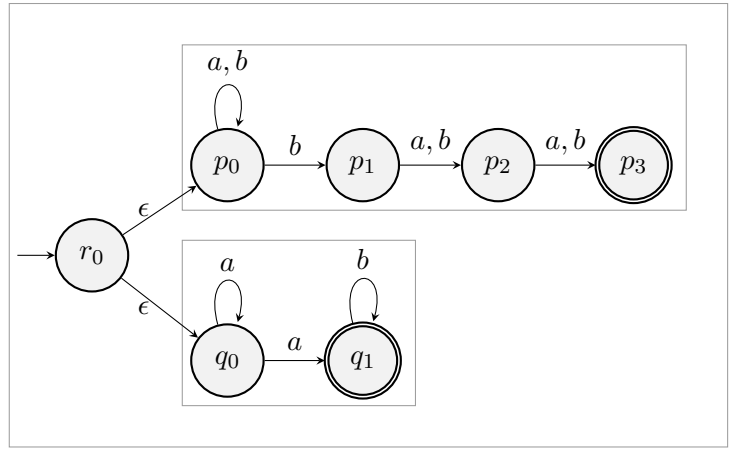

Treating $N_1$ and $N_2$ as high-level computations (by just considering their respective initial and final states), enable us to compose them into more complex automata via the operators described in Definitions \ref{def:con} and \ref{def:par}. Particularly, $N_1$ and $N_2$ can be composed into the concatenative NFA ${N_1 \oplus N_2}$ to accept strings in the language ${\{wa^mb^n\}}$. The structure of such a composite is depicted in Figure \ref{fig:n1-con-n2}. Figure \ref{fig:n1-par-n2} shows that $N_1$ and $N_2$ can also be composed into the parallel NFA $N_1 \otimes N_2$ to recognising the language ${L_1 \cup L_2}$. As it is well-known that regular languages are closed under concatenation and union, both ${N_1 \oplus N_2}$ and ${N_1 \otimes N_2}$ can in turn be inductively composed into more complex NFAs using the same composition operators. That is, our MHC satisfies closure. It also satisfies the closed-world property since it is well-known that every (classical) NFA does not admit external data streams while computing.

Control flow between composed NFAs is implicitly defined in the added state transitions of a composite NFA. For instance, the transition ${\delta(p_3,\epsilon)=\{q_0\}}$ of ${N_1\oplus N_2}$ serves to implicitly pass control from $N_1$ to $N_2$, whereas the transition ${\delta(r_0,\epsilon)=\{p_0,q_0\}}$ of ${N_1 \otimes N_2}$ defines implicit control for the simultaneous activation of $N_1$ and $N_2$. 

Although data flow is not directly supported by our MHC, it is important to analyse how data is processed within a composite automaton. In the case of a concatenative NFA, an input string is esentially computed in two different chunks. In a parallel NFA, there is no data partitioning but each composed NFA computes the whole input independently. For instance, if $aabaaaab$ needs to be processed by ${N_1\otimes N_2}$ then both $N_1$ and $N_2$ would compute and reject on $aabaaaab$ simultaneously. Injecting the same input into ${N_1\oplus N_2}$ would lead to $N_1$ passing control to $N_2$ just after implicitly determining that $aabaa$ is in $L_1$. After receiving control, $N_2$ would just compute and accept on $aab$. If we were to extend our MHC with the notion of explicit data flows (e.g., by introducing direct message passing), ${N_1\oplus N_2}$ could sequentially pass $aab$ from $N_1$ to $N_2$. By the same token, we could also enable data replication within ${N_1\otimes N_2}$ to make sure that both $N_1$ and $N_2$ receive a copy of the whole input before computing.

\section{What Classes of MHCs Do Currently Exist?}
\label{sec:classification}

In this section, we classify MHCs into three different classes: state-oriented, data-oriented and control-oriented. For each of them, we provide a brief description and present a few examples.

\begin{sideBar}
We devise three major classes of MHCs, namely state-oriented, data-oriented and control-oriented.
\end{sideBar}

\emph{State-oriented MHCs} are built upon traditional Finite State Machines (FSMs) to enable the description of complex computing systems. The most representative models in this class are the so-called \emph{Communicating Finite State Machines} (CFSMs) \cite{brand_communicating_1983} and \emph{Hierarchical Finite State Machines} (HFSMs) \cite{harel_statecharts_1987}.\footnote{Note that there also is a hybrid variant called \emph{Communicating Hierarchical State Machines} \cite{alur_communicating_1999}.} CFSMs introduce ``send" and ``receive" operations in their semantics to enable concurrent interactions among distinct FSMs via (unbounded) First-In-First-Out (FIFO) channels. The overall behaviour of a system is described through traces of configurations each of which is a tuple of both the state of the FSMs involved and the content of FIFO channels. Extensions of CFSMs include open CFSMs \cite{barbanera_connecting_2019} and parameterised CFSMs \cite{bollig_logic_2014}. The other subclass of state-oriented MHCs is that of HFSMs which allow the specification of nested FSMs in the form of so-called superstates. Due to their simplicity to modelling complex high-level computations, HFSMs have influenced the design of several industry-oriented formalisms such as the UML Superstructure specification which, in turn, has become the \emph{de facto} standard for modelling complex software systems. In Section \ref{sec:example}, we designed a simple MHC which can be classified as a HFSM. Other examples pertaining to this class are Hierarchical Featured State Machines \cite{fragal_hierarchical_2019} and scope-dependent HFSMs \cite{la_torre_verification_2008}.

\emph{Data-oriented MHCs} originated in late sixtees as an attempt to model concurrent computations in the form of direct message passing among a collection of interacting computing devices (usually referred to as actors). In these models, control flow is not explicit but implicit in the collaborative exchange of data. Here, data-driven computations are typically expressed as directed graphs in which nodes and edges denote data processing actors and explicit communication channels for directionally exchanging data streams, respectively \cite{dennis_first_1974}.\footnote{Some data-oriented MHCs (e.g., Token Flows \cite{buck_scheduling_1993}) introduce special actors to enable control constructs such as branching.} Perhaps the most representative MHCs pertaining to this class are \emph{Kahn Process Networks} (KPNs) \cite{kahn_semantics_1974} and the \emph{actor model} \cite{hewitt_viewing_1977}. On the one hand, KPNs define denotational semantics for modelling (high-level) concurrent actors that communicate through FIFO queues. Although operational semantics have been proposed to explicitly describe the flow of data values \cite{dennis_first_1974}, it has been showed that such a semantics coincide with the original denotational notion of KPNs \cite{plotkin_semantics_2009}. The operational idea is that there is a static network of actors in which every actor consumes input tokens, perfoms some computation on those tokens and produces output tokens that can potentially be consumed by other actors. The actor model does not operate on a static network of data processing actors, but it allows actors to dynamically create other actors, send messages to other actors and decide how to handle data. Other data-oriented MHCs that have been proposed over time (many of them variants of KPNs) include Synchronous Data Flow \cite{lee_synchronous_1987}, dataflow process networks \cite{lee_dataflow_1995}, reactive process networks \cite{geilen_reactive_2004} and synchronous blocks \cite{edwards_semantics_2003}. Interestingly, there are some MHCs (e.g., Reo \cite{arbab_reo_2004}) which do not allow direct data passing among computing devices, but data exchanges are ``coordinated exogenously" via well-defined algebraic dataflow structures.

\emph{Control-oriented MHCs} enable the specification of explicit order in which individual computing devices need to be invoked so that interacting devices behave as passive units of computation. Examples of MHCs belonging to this class are \emph{exogenous connectors for encapsulated components} \cite{lau_software_2006}, \emph{Behaviour Trees} \cite{colledanchise_behavior_2018} and \emph{workflow control flow models} \cite{russell_workflow_2016}. Particularly, exogenous connectors allow the formation of hierarchical control flow structures that define (exogenous) coordination for the invocation of different computing devices. A few extensions to this model have been proposed over time (e.g., \cite{lau_control_2009,arellanes_decentralized_2023}). Behaviour trees are similar to exogenous connectors in the sense that hierarchical control flow structures are formed. The difference lies in the operational semantics of control flow coordination. Over time, several extensions and enhacements of Behaviour Trees have been proposed (see \cite{iovino_survey_2022}), mainly to modelling modular Robot behaviour. Apart from exogenous connectors and Behaviour Trees, we also have workflow control flow models which define formal rules to govern the computation of workflows. Here, control flow specifies the order in which workflow activities are activated. A workflow activity is a fundamental unit of computation which can either be indivisible or contain other activities. Specific examples of workflow control flow models include  Workflow Nets \cite{van_der_aalst_application_1998}, formal BPEL process models \cite{ouyang_formal_2007}, YAWL \cite{van_der_aalst_yawl_2005}, among others.\footnote{A workflow engine is an abstract machine able to receive a workflow control flow model $M$ and a workflow specification $S$, before computing the activities in $S$ according to the rules imposed by $M$.}  

It is important to note that the three classes of MHCs we just considered are not exhaustive but indicative of the vast range of MHCs that have been proposed over time in the need of describing complex computations. For instance, we did not consider \emph{process calculi} \cite{baeten_process_2009} which are KPN-influenced models that borrow properties from both data- and control-oriented MHCs, in order to enable the description of inter-process communication through well-defined algebraic laws for equational reasoning. In the last years, there has been an increasing tendency to move towards describing high-level processes via \emph{algebras over operads of wiring diagrams} \cite{yau_operads_2018} which provide formal constructs to reasoning about functional and concurrent computations in an intuitive yet rigorous manner, typically in the language of symmetric monoidal categories wherein morphisms express high-level computations that can graphically be depicted as string diagrams \cite{piedeleu_introduction_2023}.\footnote{String diagrams have their roots in the Penrose graphical notation for tensor networks \cite{penrose_applications_1971}.} We believe that such category-theoretic operadic models are collectively forging the foundations of promising compositional MHCs. Examples within this paradigm include the calculus of signal flows \cite{bonchi_calculus_2017}, the resource calculus \cite{bonchi_diagrammatic_2019} and the zx-calculus \cite{coecke_interacting_2011}. Although not operadic in nature, other formalisms deserving attention are the so-called \emph{component models} \cite{lau_introduction_2017} since many of them form MHCs themselves. For instance, Reo \cite{arbab_reo_2004} and exogenous connectors \cite{lau_software_2006} are component models that yield data- and control-oriented MHCs, respectively, as mentioned previously.

\section{Conclusions and Future Directions}
\label{sec:conclusions}

In this paper, we introduced a new class of models of computation referred to as MHCs. Contrary to their low-level counterpart, the purpose of a MHC is to define interactions among diverse computing devices through a certain composition mechanism, so that interaction results from composition (not the other way round). As interactions occur outside the internal structure of each composed computing device, a MHC separates computation from interaction and induces modularity by treating computing devices as black bloxes. Composition can be done either algebraically or non-algebraically in order to specify (explicit/implicit) control flow and, optionally, (explicit/implicit) data flow. 

In Section \ref{sec:example}, we presented a closed MHC for facilitating the interaction of NFAs through an algebraic composition mechanism that defines implicit control flow only. Through this example, we demonstrated the modularity property that permeates the notion of MHCs. Although our examplary MHC satisfies closure over NFAs, there are other MHCs that do not necessarily compose state machines so as to deal with the well-known state explosion problem, e.g., \cite{arbab_reo_2004,lau_software_2006}. There even are MHCs that provide mechanisms to sequentially compose Turing Machines \cite{goldin_turing_2004}. In Section \ref{sec:classification}, we classified MHCs into three major classes: state-oriented, data-oriented and control-oriented. Although our review is introductory only, it can serve as a starting point to devise a more complete classification scheme (or even a whole taxonomy). For a more detailed analysis and comparison of MHC classes, one can consider the properties presented in Section \ref{sec:mhc} and beyond such as determinism vs non-determinism or algebraic versus non-algebraic.

We believe that algebraically composing computing devices is of paramount importance to tame the complexity of interactions, especially as the size of computing devices becomes larger and larger. Algebraic composition can also be beneficial to compositionally verify certain properties such as termination and reachability. Nevertheless, there are number of directions that need to be addressed before unleashing the full potential of MHCs, including (i) compositional semantics for concurrent high-level computations, (ii) design of programming languages built upon the notion of MHCs and (iii) classification of expressive power (especially of open MHCs), just to name a few directions. 

Moving up the ladder of abstraction from low- to high-level computations resembles the paradigm shift from low- to high-level programming languages. It is also analogous to the change in perspective from concrete mathematical structures to high-level ones (as in Category Theory). The paradigm shift we present in this paper clearly indicates that raising the level of abstraction is inevitably fundamental to deal with the intrinsic complexity that surround us. Accordingly, we envision that MHCs will play a crucial role in modelling complex, large-scale computing systems (or systems of systems) across various domains in the coming years. We invite the theoretical computing community to continue exploring and expanding the promising and intriguing frontiers of what we call Models of High-Level Computation. 

\bibliographystyle{unsrt}
\bibliography{refs}
\end{document}